\documentclass{jpp}
\usepackage{graphicx}
\usepackage{epstopdf, epsfig}
\usepackage{hyperref}
\usepackage{amsmath,amsfonts,amssymb}
\usepackage{subfigure}

\newcommand{\pa}[2]{\frac{\partial #1}{\partial #2}}
\newcommand{\paf}[2]{\partial #1/\partial #2}
\newcommand{\ve}[1]{\mathbf{#1}}

\graphicspath{{figs}}

\shorttitle{Bounce-Averaged Theory In Multi-Well Plasmas}
\shortauthor{I. E. Ochs}

\title{Bounce-Averaged Theory In Arbitrary Multi-Well Plasmas: Solution Domains and the Graph Structure of their Connections}

\author{I. E. Ochs\aff{1}\corresp{\email{iochs@princeton.edu}}}

\affiliation{\aff{1}Department of Astrophysical Sciences, Princeton University, Princeton, NJ, USA}

\begin{document}

\maketitle

\begin{abstract}
Bounce-averaged theories provide a framework for simulating relatively slow processes, such as collisional transport and quasilinear diffusion, by averaging these processes over the fast periodic motions of a particle on a closed orbit.
This procedure dramatically increases the characteristic timescale and reduces the dimensionality of the modeled system.
The natural coordinates for such calculations are the constants of motion (COM) of the fast particle motion, which by definition do not change during an orbit.
However, for sufficiently complicated fields---particularly in the presence of local maxima of the electric potential and magnetic field---the COM are not sufficient to specify the particle trajectory.
In such cases, multiple domains in COM space must be used to solve the problem, with boundary conditions enforced between the domains to ensure continuity and particle conservation.
Previously, these domains have been imposed by hand, or by recognizing local maxima in the fields, limiting the flexibility of bounce-averaged simulations.
Here, we present a general set of conditions for identifying consistent domains and the boundary condition connections between the domains, allowing the application of bounce-averaged theories in arbitrarily complicated and dynamically-evolving electromagnetic field geometries. 
We also show how the connections between the domains can be represented by a directed graph, which can help to succinctly represent the trajectory bifurcation structure.
\end{abstract}


\section{Introduction} 

In a magnetic confinement device such as those used for fusion reactors, particles typically experience fast helical motion along magnetic field lines.
These motions then form closed periodic orbits, sometimes with precession around an angular coordinate.
For modeling many macroscopic phenomena that occur on relatively slow timescales compared to the orbit timescale, it is much more efficient to average the effect of the process over the orbit, rather than to solve the full equations of motion for the orbit, as this involves a much slower-timescale calculation in fewer dimensions.
This simplification is the basis for bounce-averaged Fokker Planck (BAFP) theory, which can be used to model collisional transport \citep{Rosenbluth1957FokkerPlanckEquation,Hager2016FullyNonlinear}, quasilinear wave-particle interactions \citep{Stix1975FastwaveHeating,Bernstein1981RelativisticTheory,Fisch1987TheoryCurrent,Eriksson1994MonteCarlo,Herrmann1997CoolingEnergetic,Herrmann1998CoolingAlpha,Fisch1992InteractionEnergetic}, and radiation emission \citep{Bilbao2023RadiationReaction,Zhdankin2023SynchrotronFirehose} and absorption \citep{Ochs2024ElectronTail,Ochs2024SynchrotrondrivenInstabilities} in tokamak \citep{Harvey1992CQL3DFokkerplanck}, stellarator \citep{Mynick1986BounceaveragedFokkerPlanck,Nemov1999Evaluation1,Velasco2020KNOSOSFast,dHerbemont2022FiniteOrbit}, and mirror \citep{BenDaniel1962ScatteringLoss,Marx1970EffectsSpatial,Matsuda1986RelativisticMultiregion,Egedal2022FusionBeam,Frank2024IntegratedModelling} plasmas.

Each closed orbit is associated with conserved constants of motion (COM) in a lower dimensional space.
For instance, the energy $\epsilon$, magnetic moment $\mu$, and azimuthal momentum $p_\theta$ form the typical COM for a steady-state axisymmetric magnetic field arrangement such as a tokamak or magnetic mirror.
Since they are constant on each averaged trajectory, the COM represent a natural coordinate system for solving BAFP problems.
Thus, the first step is often to express the distribution function as a function only of the constants of motion $\ve{Z}$, averaging over trajectories that differ only by the phase of the motion.

\begin{figure}
	\centering
	\includegraphics[width=0.6\linewidth]{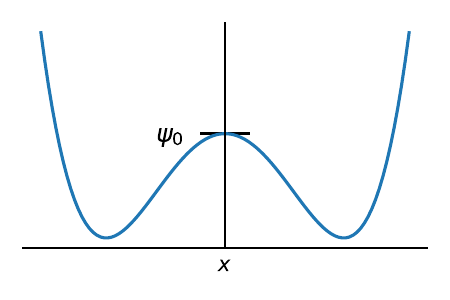}
	\caption{A one-dimension double well scalar potential $\psi(x)$. 
		The energy $\epsilon$ uniquely defines a single trajectory for $\epsilon>\psi_0$.
		However, at $\epsilon<\psi_0$, there are two trajectories that have the same $\epsilon$, corresponding to trapping in the two wells.
		Thus, the function $f(\epsilon)$ is not necessarily well-defined below $\epsilon=\psi_0$.}
	\label{fig:DoubleWell1D}
\end{figure}

This procedure often works well for simple field arrangements without local maxima in the fields.
However, it runs into a problem if there are local maxima.
Consider, for instance, particles in the 1-dimensional double well potential $\psi(x)$ shown in Fig.~\ref{fig:DoubleWell1D}.
This 1D arrangement (with its associated 2D phase space) has the constant of motion $\epsilon = m v^2/2 + \psi(x)$.
For particles with energy $\epsilon > \psi_0$, the value of $\epsilon$ defines a single trajectory shape that traverses both positive and negative $x$, and there is no problem defining $f(x,v) \rightarrow f(\epsilon)$ to average slow processes over the bounce motion.
However, for $\epsilon < \psi_0$, the value of $\epsilon$ no longer contains complete information about the trajectory, which bifurcates into two trajectories: one trapped in the well at $x > 0$, and the other trapped in the well at $x < 0$.
Since the occupancy of each well can, in general, be different, the function $f(\epsilon)$ is no longer well-defined. 

Multiple wells, however, do not mean that BAFP cannot be used to model multi-well configurations.
The resolution lies in splitting the space into three domains, representing (i) the passing trajectories at $\epsilon>\psi_0$, (ii) the trajectories trapped in the left well, and (iii) the trajectories trapped in the right well.
Boundary conditions \citep{Matsuda1986RelativisticMultiregion,dHerbemont2022FiniteOrbit} then must be enforced between the domains to ensure continuity of the distribution function and particle conservation.

These bifurcations of trajectories are common in mirror physics, and the solution of their associated BAFP problems appeared several times in the study of tandem mirrors \citep{Cohen1980ParticleEnergy,Matsuda1986RelativisticMultiregion,Katanuma1986ThermalBarrier,Katanuma1987FokkerPlanck,Fowler2017NewSimpler}, which are characterized by internal peaks in both the magnetic field and electric potential around the tandem end plugs.
However, in solving these problems, the boundary conditions were typically imposed by hand for very specifically shaped fields, and were only ever attemped in two dimensions.
In moving to 3D computational modeling of modern mirrors, with fields that can be significantly more complicated due to sloshing ion distributions \citep{Egedal2022FusionBeam,Endrizzi2023PhysicsBasis,Frank2024IntegratedModelling}, multi-mirror configurations \citep{Beery2018PlasmaConfinement,Miller2021RateEquations}, ponderomotive plugs \citep{Miller2023RFPlugging,Rubin2023MagnetostaticPonderomotive,Ochs2023CriticalRole,Rubin2024FlowingPlasma,Kolmes2024CoriolisForces,Rubin2025PonderomotiveBarriers,Ochs2025PreventingAsh}, or centrifugal forces \citep{Bekhtenev1980ProblemsThermonuclear,Cho2005ObservationEffects,Schwartz2024MCTrans0D,Kolmes2025IonMix}, determining the domains by hand at the beginning of the simulation may be difficult or impossible. 
Furthermore, while modern codes for neoclassical transport in stellarators \citep{Nemov1999Evaluation1,Velasco2020KNOSOSFast} solve similar multi-well matching problems, they generally assume that the electrostatic potential is constant enough on a trajectory to not influence the trapping condition, which is distinctly untrue for mirror plasmas.
Thus, it is important to formalize conditions and methods for establishing solution domains for arbitrarily complicated and arbitrarily high-dimensional multi-well BAFP problems, with wells that result from variations in both the magnetic field and electric potential.
Establishing these conditions, particularly for mirror-like plasmas, is the goal of this paper.

To fulfill of this goal, we begin in Sec.~\ref{sec:BAFP}, by reviewing the basics of BAFP theory. 
In Sec.~\ref{sec:MultipleDomainIntro}, we show how trajectories bifurcate for common mirror configurations, establishing notation and motivating the need for a general method to identify consistent domains.
In Sec.~\ref{sec:DomainRules}, we formalize the requirements for establishing cosistent domains, and show how to fully partition COM space $\ve{Z}$ so that the distribution $f(\ve{Z})$ is single-valued and well-defined.
In Sec.~\ref{sec:BoundaryConditions}, we establish the boundary conditions between these domains, allowing BAFP to be solved in the global space despite an arbitary number of trajectory bifurcations. 
Because these boundary conditions can be complicated, with different domains matched at different parts of the boundary, in Sec.~\ref{sec:Graphs} we show how the connections between the various domains can be simply visualized via a directed graph.
Finally, in Sec.~\ref{sec:Conclusion}, we discuss how these ideas enable the design of simulations for dynamically-evolving multi-well plasma systems.

\section{Bounce-Averaged Theory in COM Coordinates} \label{sec:BAFP}

%

Here, we briefly review the basics of bounce-averaged Fokker-Planck (BAFP) theory.
We move quickly, with a primary focus on magnetic mirror geometries.


In a magnetic mirror, whether simple, tandem, or centrifugal, particles in the mirror live on trapped trajectories.
On the shortest timescale, the gyroperiod $1/\Omega_c$, the particle gyrates around the magnetic field line.
On the next longest timescale, the bounce period $\tau_b \sim L_\parallel / v_\parallel$, the particle transits the mirror, completing a closed orbit.

Absent instabilities, particle deconfinement in such mirrors occurs due to collisional processes.
Thus, on a collision timescale $\tau_c \gg \tau_b, 1/\Omega_c$, collisions push the particle onto new orbits.
Eventually, the particle is pushed onto an orbit that leaves the device.

Obviously, it is inefficient (whether analytically or computationally) to resolve the gyroperiod and bounce timescales and complex trajectories when evaluating the slower diffusion processes that drive particles out of the device.
Suitably averaging over these fast timescales to produce a lower-dimensional set of equations describing these slow processes is the goal of BAFP.

The first step in deriving the theory is to transform to coordinates that are constant along a trajectory.
For an axisymmetric, nonrelativistic plasma, a suitable set of constants of motion (COM) are:
\begin{align}
	\mu &\equiv \frac{1}{2} \frac{m v_\perp^2}{B(\ve{x})} \label{eq:mu}\\
	\epsilon &\equiv \frac{1}{2} m v_\parallel^2 + \mu B(\ve{x}) + \psi (\ve{x}) \label{eq:epsilon}\\
	p_\theta &\equiv m \Omega_r r^2 + q \Phi (\ve{x}).
\end{align}
Here, $\mu$ is the magnetic moment of the particle, $\epsilon$ is the energy, and $p_\theta$ is the azimuthal momentum.
Additionally, $v_\parallel$ and $v_\perp$ are the velocity components parallel and perpendicular to the magnetic field, $\Omega_r(\Phi)$ is the rotation frequency of the magnetic surface, and $q$ and $m$ are the charge and mass of the particle.
The potential energy term $\psi$ is given by:
\begin{align}
	\psi = q \phi_\parallel + \frac{1}{2} m r^2 \Omega_r^2
\end{align}
where $\phi_\parallel$ is a potential coordinate along the flux surface \citep{Kolmes2024MassiveLonglived}.
Finally, $\Phi$ is the flux function, given by:
\begin{align}
	\Phi = r A_\theta.
\end{align}

In looking at $p_\theta$, note that the ratio of the first term to the second is given by:
\begin{align}
	\frac{m \Omega_r }{q A_\theta} \sim \frac{\Omega_r}{\Omega_c}, 
\end{align}
i.e. $p_\theta$ conservation implies that particles stay on their flux surface for particles far from the Brillouin limit \citep{Rax2015BreakdownBrillouin}.
We assume in the following that we are in this regime, taking $\Phi$ as the third COM in place of $p_\theta$.

To reach our six dimensions of phase space, we need three other variables.
These we take to be the azimuthal angle $\theta$, the gyro-angle $\alpha$, and the distance along a field line $s$.
These variables $(\theta,\alpha,s)$ are the variables we will integrate out to form the bounce-averaged theory.
Integrating out $\alpha$ is the basis for gyro or drift kinetics; integrating out $\theta$ makes the theory axisymmetric, and integrating out $s$ results in the bounce average.

Now, in general, we will start with an equation of the form:
\begin{align}
	\sqrt{g_X} \pa{f}{t} = \pa{}{X^i} \left[ \sqrt{g_X} \, \Gamma_X^i (f,\ve{X})\right] + \sqrt{g_X} S(\ve{X}) ,
\end{align}
where $\Gamma^i(f,\ve{X})$ is a flux operator, $S(\ve{X})$ is a source operator, and $\sqrt{g_X}$ is the volume element in the space $\ve{X} \equiv (\ve{x},\ve{p})$.
This equation is in conservation form; integrating the left hand side gives the total change in particle number in a given region, while the first term on the RHS reduces to a surface integral of the flux out of the region.

To get our bounce-averaged Fokker-Planck equation, we first need to convert to the 6D space $\ve{Y}$ that contains the COMs, i.e. $\ve{Y} \equiv (\epsilon,\mu,\Phi,\theta,\alpha,s)$ .
Because the metric in phase space $\ve{X} \equiv (\ve{x},\ve{v})$ is $g_{ij} = \delta_{ij}$, corresponding to a unit volume element, the volume element $\ve{dV} =\sqrt{g_Y} \ve{dY}$ in the new coordinates is:
\begin{align}
	\sqrt{g_Y} = 2\sqrt{\left| \pa{X^m}{Y^i} \pa{X^n}{Y^j} \delta_{ij} \right| } = \frac{\sqrt{2}}{m^{3/2}\sqrt{\epsilon - \mu B - \psi}}. 
\end{align}
Here, the factor of 2 comes from the fact that positive and negative $v_\parallel$ are condensed into the same portion of COM space.
We also convert the flux $\Gamma$ to $Y$ space. 
For instance, for a Fokker-Planck flux operator:
\begin{align}
	\Gamma^i_X = A_X^i f + D_X^{ij} \pa{f}{X^j},
\end{align}
we take 
\begin{align}
	\Gamma^i_Y = \pa{Y^i}{X^m} A^m f + \pa{Y^i}{X^m} \pa{Y^j}{X^n}D^{mn} \pa{f}{Y^j},
\end{align}

Now, we assume that $f$ is independent of $(\theta,\alpha,s)$, i.e. take $f(Z)$, $\ve{Z} \equiv (\epsilon,\mu,\Phi)$.
Then, we integrate the Fokker-Planck equation over the non-COM coordinates in $\ve{Y}$.
The integrals over $\theta$ and $\alpha$ are trivial, leading to two factors of $2\pi$.
The integral over $s$ is more nontrivial, but results in:
\begin{align}
	\sqrt{g_Z} \pa{f}{t} = \pa{}{Z^i} \left( \sqrt{g_Z} \, \Gamma^i (f,\ve{Z})\right) + \sqrt{g_Z} S_Z(\ve{Z}) ,
\end{align}
where
\begin{align}
	\sqrt{g_Z} &\equiv 4 \pi^2 \int_{s_1}^{s_2} ds \sqrt{g_Y} \\
	\Gamma_Z^i &\equiv \frac{1}{\sqrt{g_Z}} \int_{s_1}^{s_2} ds \sqrt{g_Y} \, \Gamma^i_Y \\
	S_Z &\equiv \frac{1}{\sqrt{g_Z}} \int_{s_1}^{s_2} ds \sqrt{g_Y} S_Y ,
\end{align}
and where $s_1$ and $s_2$ are the two outer limits of the orbit.

Noting that $v_\parallel(s) = \sqrt{2 (\epsilon - \mu B(s) - \psi(s) )/ \mu}$, we see that the volume element $\sqrt{g_Z}$ is proportional to the total bounce time $\tau_b = \int_{s_1}^{s_2} ds/v_\parallel(s)$, while the operators $\Gamma_Z^i$ and $S_Z^i$ are averaged based on the time $ds/v_\parallel(s)$ they spend in each region.
Thus, the phase-space volume average and temporal bounce average are equivalent.




\section{Multiple Wells and the Need for Multiple Domains} \label{sec:MultipleDomainIntro}

Bounce-averaged theory provides a massive simplification to the collisional Vlasov equations.
However, it relies on one major assumption: that $f$ can be written as a function of the COM coordinates alone: $f = f(\ve{Z})$.
As we will see, this condition is often violated in scenarios relevant for modern mirrors.
When it is violated, a single function $f(\ve{Z})$ on COM space is no longer adequate to describe the complete system.
Then, the problem must be solved on a set of bounded domains, with appropriate boundary conditions enforced on the shared boundaries between domains.

To explore these issues, we will make two simplifications to the theory above.
First, we will look at equations for a single value of $\Phi$, looking only at the 2D COM space of $(\epsilon,\mu)$.
Despite this reduction in dimension, we note that the methods presented here generalize straightforwardly to 3D COM space.

Second, we will consider a piecewise-constant series of potentials $\psi(n(s))$ and magnetic fields $B(n(s))$, with the $n$th piece of width $\Delta s_n$, so that the integrals in the bounce-averaged theory become sums.
Of course, this is equivalent in the limit of $\Delta s \rightarrow 0$ and the number of partitions going to infinity, but working with a small discrete number of areas will allow us to develop the discussion much more clearly.

Consider, then, particle dynamics along a field line.
As is often conventional, we will take the mirror dynamics to be symmetric about the mirror midplane, and so consider a sequence of $\psi(n)$, $B(n)$, with $n \in \{0,...,N-1\}$, and with $n=0$ corresponding to the midplane and $n = N-1$ corresponding to the mirror throat.
We normalize the potentials and fields to their midplane values, with $\psi(0) = 0$ and $B(0) = 1$, and arbitrary units for $\psi$ and $\epsilon$.

Given sequences $\psi(n)$ and $B(n)$, we can determine the particles allowed in any given field line segment $n$.
For a particle to be allowed in a region, it must have positive parallel kinetic energy, $v_\parallel^2 > 0$.
From Eq.~(\ref{eq:epsilon}), we see that this requires:
\begin{align}
	\epsilon \geq \mu B(n) + \psi(n) \label{eq:AllowedRegion}.
\end{align}
Each pair $(\psi(n)$, $B(n))$ thus determines an allowed region of $(\epsilon,\mu)$ space, i.e. the region satisfying Eq.~(\ref{eq:AllowedRegion}).

\begin{figure}
	\centering
	\includegraphics[width=0.8\linewidth]{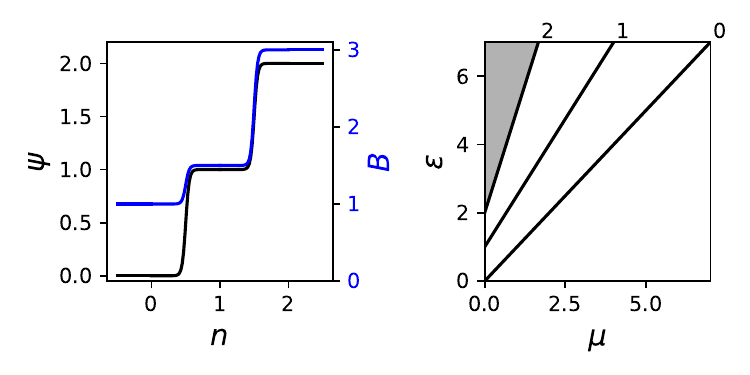}
	\caption{Discretized magnetic field $B$ and potential energy $\psi$ as a function of axial segment $n$ (left) and COM-space acessibility plot (right) for electrons in a magnetic mirror with a (typical) outward-pointing electric field.
		In the COM-space accessibility plot, each line $n\in \{1,2,3\}$ represents the boundary below which particles do not have enough kinetic energy to enter that axial segment.}
	\label{fig:COMPlot_SimpleMirrorElectrons}
\end{figure}

The simplest case is provided by electrons in a simple mirror (which also applies to both electrons and ions in a centrifugal mirror).
An ambipolar potential generally forms \citep{Pastukhov1974CollisionalLosses,Najmabadi1984CollisionalEnd,Ochs2023ConfinementTime}, confining electrons and repelling ions.
Thus, for electrons, both $\psi$ and $B$ increase as a function of line segment index $n$, and so particles are best trapped at the midplane (Fig.~\ref{fig:COMPlot_SimpleMirrorElectrons}).
This is the ideal situation for BAFP theory: trajectories that can access segment $n=1$ also access segment $n=0$, so the distribution function $f(\epsilon,\mu)$ is clearly single-valued.
To apply a BAFP theory, operator averages are performed over segment $n = 0$ for the particles above line 0 and below line 1, and over segments $n\in \{0,1\}$ for the particles that are above line 1 and below line 2.
A similar situation will occur whenever $\paf{\psi}{s} \geq 0$ and $\paf{B}{s} \geq 0$ for the entire field line from midplane to throat.
Such a plasma can even be modeled using midplane momentum coordinates, since every allowed trapped trajectory reaches the midplane ($n=0$).

\begin{figure}
	\centering
	\includegraphics[width=0.8\linewidth]{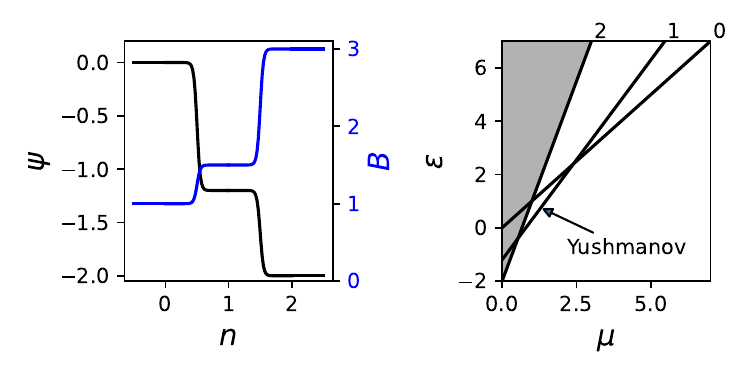}
	\caption{Discretized magnetic field $B$ and potential energy $\psi$ as a function of axial segment $n$ (left) and COM-space acessibility plot (right) for ions in a magnetic mirror with a (typical) outward-pointing electric field.
		Because of the decreasing electric potential toward the edge of the device, some ions get trapped between the mirror throat and the midplane, i.e. at $n=1$.
		These ions are referred to as ``Yushmanov-trapped.''}
	\label{fig:COMPlot_SimpleMirrorIons}
\end{figure}

A somewhat more complex case is provided by ions in a simple mirror.
Because the potential is repulsive, $\psi$ now decreases as a function of $n$ (Fig.~\ref{fig:COMPlot_SimpleMirrorIons}).
Thus, a class of ions can develop which are trapped off the midplane in the potential well, despite the higher magnetic field there.
These are known as ``Yushmanov-trapped'' particles \citep{Post1987MagneticMirror,Yushmanov1966ConfinementSlow}.
In a plasma with Yushmanov-trapped particles, a midplane-momentum coordinate theory becomes insufficient, because these particles exist in the plasma but do not reach the midplane.
However, the distribution function $f(\ve{Z})$ is still clearly single-valued, so BAFP theory can still be used, solved on a single domain in COM space.
The operator averages are performed over $n=1$ for the Yushmanov particles, over $n=0$ for the particles above line 0 and below line 1, and over $n\in\{0,1\}$ for the passing particles that traverse both segments.

\begin{figure}
	\centering
	\includegraphics[width=0.8\linewidth]{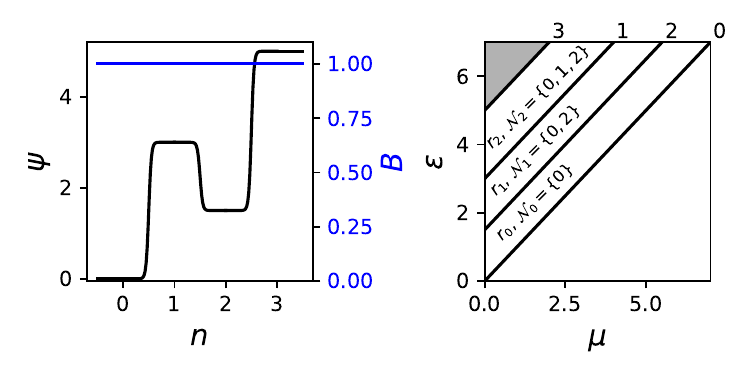}
	\caption{Discretized magnetic field $B$ and potential energy $\psi$ as a function of axial segment $n$ (left) and COM-space acessibility plot (right) for a scenario with a constant magnetic field and and internal potential maximum.
		This scenario exhibits a bifurcation of trajectories around $n=1$, and thus the bounce-averaged theory for this space requires 3 solution domains.}
	\label{fig:COMPlot_PotentialMax}
\end{figure}

As the magnetic geometry and potentials become more complex, the single-valuedness can quickly break down.
For instance, consider a case where the magnetic field is constant, but the potential has an intermediate peak (Fig.~\ref{fig:COMPlot_PotentialMax}).
Now, as in the 1D double well in Fig.~\ref{fig:DoubleWell1D}, $\epsilon$ and $\mu$ are not sufficient to define a particle trajectory, since a particle with insufficient $\epsilon$ to traverse the peak can be trapped on one side or the other.
For instance, in Fig.~\ref{fig:COMPlot_PotentialMax}, in region $r_1$ where $(\epsilon,\mu)$ is below line 1 and above line 2, the particle can either be on a trajectory that is trapped in segment $n=0$, or a trajectory trapped in segment $n=2$.
There is no reason for particles on these different trajectories to share a value for the distribution function; thus, $f(\epsilon,\mu)$ fails to be single-valued in this region.
[A similar bifurcation occurs for a local maximum of the magnetic field; the only difference is that changing $B$ would change the slope of the lines in $(\epsilon,\mu)$ space, rather than their $\epsilon$-intercept.]

When such a bifurcation of the trajectory happens, the domain on which the BAFP equations are solved must be split, to keep $f(\epsilon,\mu)$ single-valued on each domain.
Each domain consists of (a) an area in $(\epsilon,\mu)$ space, and (b) a continuous subset of the field line segments $n$ over which the bounce-average is taken.
For the case in Fig.~\ref{fig:COMPlot_PotentialMax}, three domains are required: the domain $d_1$ consisting of region $r_2$, bounce-averaged over field line segments $n \in \{0,1,2\}$; the domain $d_2$ consisting of regions $r_0$ and $r_1$, bounce-averaged over field line segments $n \in \{0\}$; and the domain $d_3$ consisting of region $r_1$, bounce-averaged over field line segments $n \in \{2\}$. 

To solve the problem, the domains must be stitched together.
This requires two boundary conditions at the boundary (line 1).
First, $f(\epsilon,\mu)$ must be continuous at the boundary.
Second, the phase-space flux must be continuous, i.e.
\begin{align}
	n_i \Gamma_Z^i \bigr|_{d_1} = n_i \Gamma_Z^i \bigr|_{d_2} + n_i \Gamma_Z^i \bigr|_{d_3}, \label{eq:BoundaryMatching}
\end{align}
where $n_i$ is a shared normal vector to the surface in $(\epsilon,\mu)$ space.

\begin{figure}
	\centering
	\includegraphics[width=0.3\linewidth]{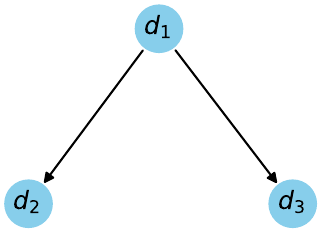}
	\caption{Directed graph structure of boundary conditions for the field configuration in Fig.~\ref{fig:COMPlot_PotentialMax}.}
	\label{fig:Graph_PotentialMax}
\end{figure}

We can see from the flux conservation equation that the two domains $d_2$ and $d_3$ merge into $d_1$.
This suggests a representation of the relationship between the domains as a directed graph, with the directionality going from the domain representing a single passing trajectory to the domains representing bifurcated trajectories (Fig.~\ref{fig:Graph_PotentialMax}).

\subsection{Ambipolar Fields and the Need for an Automatic Method} \label{sec:AmbipolarSolve}

Historically, especially in the study of tandem mirrors with thermal barriers, these domain relations have been imposed at the beginning of the problem and left fixed for the computation \citep{Matsuda1986RelativisticMultiregion}. 
However, this is likely to be insufficient for accurate simulations of modern mirrors, which often have potential profiles determined by sloshing kinetic ion distributions \citep{Egedal2022FusionBeam,Endrizzi2023PhysicsBasis,Frank2024IntegratedModelling}.
Correctly determining the resulting potentials requires solving for the electric potential in each region self-consistently over the course of the simulation so as to enforce quasineutrality, i.e. to enforce
\begin{align}
	\sum_s Z_s \int_s f_s d\ve{Z} = 0,
\end{align}
for each region \citep{Frank2024IntegratedModelling,Kolmes2025IonMix}.
As a result, it might not be known at the beginning of the simulation where the potential peak might occur, and thus over which field line segments and regions of $(\epsilon,\mu)$ to define the domains.
Thus, dynamic detection of such domains is a necessity.

\begin{figure}
	\centering
	\includegraphics[width=0.8\linewidth]{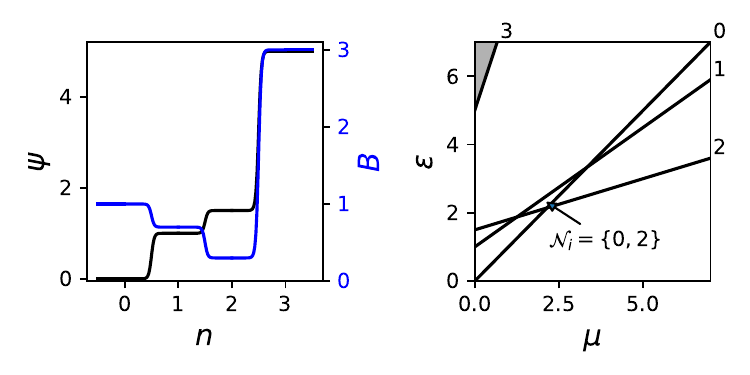}
	\caption{Discretized magnetic field $B$ and potential energy $\psi$ as a function of axial segment $n$ (left) and COM-space acessibility plot (right) for a scenario exhibiting trajectory bifurcation without a local maximum in either potential or magnetic field.
		This ``Yushmanov trajectory bifurcation'' process is discussed in more detail in Appendix~\ref{sec:TrajectoryBifurcationNonRelativistic}.}
	\label{fig:COMPlot_BifurcationNoMax}
\end{figure}

Making matters worse, the bifurcation of trajectories does not generally require such an easy-to-spot feature as a local peak in the electric or magnetic field (which is the basis for such calculations in stellarator neoclassical transport codes \citep{Nemov1999Evaluation1,Velasco2020KNOSOSFast}). 
Consider the field arrangement in Fig.~\ref{fig:COMPlot_BifurcationNoMax}.
In segment $n=1$, this has neither a local maximum in the magnetic or electric field, and thus does not seem like it would represent a barrier between local wells.
Nevertheless, as can be seen in Fig.~\ref{fig:COMPlot_BifurcationNoMax}, there is a region of $(\epsilon,\mu)$ space that can be accessed by particles in with $n \in \{0,2\}$,  but not for $n=1$.
Thus, this field setup also has a trajectory bifurcation, and must be solved using multiple grids.

The conditions required for the particular situation in Fig.~\ref{fig:COMPlot_BifurcationNoMax}, which we term a ``Yushmanov trajectory bifurcation,'' can actually be solved for (Appendix \ref{sec:TrajectoryBifurcationNonRelativistic}); it turns out that such a bifurcation will occur when
\begin{align}
	\pa{\psi}{B} < 0 \text{ and } \pa{^2\psi}{B^2} > 0. 
\end{align}
Nevertheless, in anticipating arbitrary field arrangements involving multiple wells, and particularly in looking forward to non-relativistic plasmas in 3 COM dimensions, it is clearly useful to have a generalizable method that can take the allowed regions (in COM space) for each part (in $\ve{x}$) of the plasma, and return the requisite domains and boundary conditions.
The development of such a method is the subject of the next sections.

\section{Formalizing the Requirements for Cohesive Domains} \label{sec:DomainRules}

As a starting point, we assume that we have a set of segments $\mathcal{N} = \{0,\dots, N-1\}$, with allowed areas of $(\epsilon,\mu)$ space defined for each segment $n \in \mathcal{N}$, of the form $\epsilon \geq b_n(\mu)$.
This will in general chop $(\epsilon,\mu)$ space up into a number of different regions $r_i$, with edges defined by the boundary functions $b_n(\mu)$.

We note that in formulating the problem in this way, the generalization to 3-dimensional $(\epsilon,\mu,\Phi)$ space is straightforward, since each allowed region will now be given by $\epsilon \geq b_n(\mu,\Phi)$, with the boundaries between regions given by 2-dimensional surfaces rather than 1-dimensional curves.
The following discussion thus applies immediately to more complicated and higher-dimensional problems, though to keep things clear we will restrict the discussion to nonrelativistic $(\epsilon,\mu)$ space.

Now consider a specific region $r_i$.
This region will have a set $\mathcal{N}_i \subseteq \mathcal{N}$ of the segments for which $r_i$ is an allowed region.
Ordering the set $\mathcal{N}_i$ in terms of increasing $n$, we can in general then split $\mathcal{N}_i$ into maximal continuous subsequences $\mathcal{C}_i^m$, defined such that each $\mathcal{C}_i^m$ has no breaks in the integer sequence of $n$'s.
The set of all $\mathcal{C}_i^m$ for a given region $r_i$ we denote $\boldsymbol{\mathcal{C}}_i$.
For a given $r_i$, the size (cardinality) of the set $\boldsymbol{\mathcal{C}}_i$ represents the number of distinct trajectories in that region.
For instance, returning to Fig.~\ref{fig:COMPlot_BifurcationNoMax}, the indicated region in $(\epsilon,\mu)$, which we can label as region $r_0$, has $\mathcal{N}_0 = \{0,2\}$, $\mathcal{C}_0^0 = 0$, $\mathcal{C}_0^1 = 2$, and $\boldsymbol{\mathcal{C}}_0 = \{\mathcal{C}_0^0,\mathcal{C}_0^1\} = \{\{0\},\{2\}\}$.
The fact that $|\boldsymbol{\mathcal{C}}_0| = 2$ means that there are two distinct trajectories; namely, those trapped in axial segment $n=0$, and those trapped in axial segment $n=2$.

At this point, it is helpful to define a population, denoted $p_i^m$, as the combination of the region $r_i$ (a volume in COM space) and a maximal continuous subsequence of allowed axial segments $C_i^m$ in that region.
The population defines everything other than the boundary conditions about the bounce-averaged Fokker-Planck problem in COM space: namely, it defines the region over which the problem is to be solved, and the axial segments over which the bounce average is to be taken.
Boundary conditions then come from stitching together connecting populations.

However, solving the Fokker-Planck problem for each population so defined would be extremely inefficient, since there can be many regions even for problems (such as those in Figs.~\ref{fig:COMPlot_SimpleMirrorElectrons}-\ref{fig:COMPlot_SimpleMirrorIons}) that can be solved on a single domain.
Thus, our goal will be to build up collections of compatible populations into maximal simulation domains.

Thus, define a domain $d$ as a collection of connected and pairwise-compatible populations $\{p_i^m\}$. 
To define ``connected'' and ``pairwise-compatible'', it is helpful to consider what happens as we try to extend the domain to a new population.
Thus, consider a population $p_i^m$ in domain $d$, with region $r_i$ and continuous sequence $\mathcal{C}_i^m$.
Then, try to extend the domain $d$ by adding a population with region $r_j$ bordering $r_i$.
What are the conditions that allow this to happen?

First, it only makes sense to extend the domain if there is a population on $r_j$ that contacts trajectories from $p_i^m$; i.e. if
\begin{align}
	\exists \mathcal{C}_j^n \in \boldsymbol{\mathcal{C}}_j : (r_i \text{ borders } r_j) \& (\mathcal{C}_j^n \cap \mathcal{C}_i^m \neq \{\}).
\end{align}
If this condition isn't satisfied, then trajectories in region $r_j$ are located in a spatially distinct part of the plasma from those trajectories described by $p_i^m$, and thus do not interact in a Fokker-Planck manner with the population $p_i^m$.
Thus, populations $p_i^m$ and $p_j^n$ are said to be connected if:
\begin{align}
	(r_i \text{ borders } r_j) \& (\mathcal{C}_j^n \cap \mathcal{C}_i^m \neq \{\}). \label{eq:ConnectedPopulations}
\end{align}

However, connectedness is not sufficient to ensure that the domain can be extended.
Consider, for instance, the regions $r_1$ and $r_2$ on either side of line 1 in Fig.~\ref{fig:COMPlot_PotentialMax}.
Then consider the population $p_2^0 \equiv (r_2,\{0,1,2\})$.
Going to region $r_1$, we see that there are two populations that are connected to $p_2^0$: $p_1^0 \equiv (r_1,\{0\})$, and $p_1^1 \equiv (r_1,\{2\})$.
Precisely because there are multiple connected populations in region $r_1$, the domain cannot be extended, because this multiplicity means that trajectories bifurcate at this point and a boundary condition (Eq.~\ref{eq:BoundaryMatching}) must be enforced at line 1.
Thus, in order for $p_i^m$ in region $r_i$ to be compatible with a population $p_j^n$ in region $r_j$, we must have:
\begin{align}
	\nexists \; l : (l\neq n) \& (\mathcal{C}_j^l \cap \mathcal{C}_i^m \neq \{\}). \label{eq:Compatiblity1}
\end{align}
However, this is still not sufficient, since this compatibility relation must be reciprocal; condition~(\ref{eq:Compatiblity1}) shows that $p_2^0$ is incompatible with the $p_1^0$ and $p_1^1$ in Fig.~\ref{fig:Graph_PotentialMax}, but not that $p_1^0$ and $p_1^1$ are incompatible with $p_2^0$.
Thus, in order for $p_i^m$ in region $r_i$ to be compatible with a population $p_j^n$ in region $r_j$, we must also have:
\begin{align}
	\nexists \; l : (l\neq m) \& (\mathcal{C}_i^l \cap \mathcal{C}_j^n \neq \{\}). \label{eq:Compatiblity2}
\end{align}
If, for populations $p_i^m$ and $p_j^m$, conditions~(\ref{eq:Compatiblity1}-\ref{eq:Compatiblity2}) hold, then the populations are compatible, and can exist in the same domain. 

Unfortunately, it can be shown (Appendix~\ref{sec:Nontransitivity}) that the compatibility property is not transitive.
Thus, to add a population to the domain, connectedness must only be checked for one region, but compatibility must be checked with all populations already in the domain.
This non-transitivity is why the domain was defined to be a connected and \emph{pairwise}-compatible set of populations.

To completely partition the space for solution of the Fokker-Planck problem, we must assign every population to a domain.
These domains are then connected by boundary conditions along shared boundaries.
As shown in Appendix~\ref{sec:Nontransitivity}, this decomposition is not, in general, unique.
Nevertheless, an example algorithm for performing a valid decomposition is described in Appendix~\ref{sec:Algorithm}.

At the end, each domain $d_a$ (which is a set of populations) will be associated with set of connected regions $\mathbb{R}_a$, and a single continuous set of axial segments $\mathbb{C}_a$.
Thus, for each domain $d_a$, the BAFP problem is to be performed over the area in $\ve{Z}$ that is spanned by $\mathbb{R}_a$, with the bounce average taken over the axial segments contained in $\mathbb{C}_a$.

\section{Enforcing Boundary Conditions} \label{sec:BoundaryConditions}

Now that the space is partitioned into domains, we must define boundary conditions for each domain.
As in Sec.~\ref{sec:MultipleDomainIntro}, we need to enforce continuity and particle conservation at each boundary; the key aspect will be to identify which populations must be matched.

There are two types of boundaries.
External boundaries are defined by the edge of accessible $\ve{Z}$ for the total modeled space.
These boundaries are either reflecting, if they represent the boundary of 0 kinetic energy that defines accessibility, or approximately absorbing (Dirichlet with a value of 0), if they represent the loss cone. [Though it may be noted that in more collisional mirrors \cite{Mirnov1979LinearGasdynamic,Rognlien1980TransitionPastukhov}, or mirrors with significant secondary electron emission from the walls \cite{Konkashbaev1978PossibilityDecreasing,Skovorodin2019SuppressionSecondary}, a non-zero loss cone boundary may be appropriate.]

The remaining boundaries are internal, representing matching conditions between domains.
In general, if we look at a domain $d_a$, each internal boundary can be defined by two regions; the region $r_i$ (with population $p_i^m$ in domain $d_a$), and the region $r_j$.
If $r_j$ does not represent a region within the domain $d_a$, then this represents an outer boundary of the domain $d_a$, where the boundary conditions must be applied.
We can denote this boundary by $b_{ij}^a$.

Now, note that of the two regions $r_i$ and $r_j$, one of these regions will always have a set of accessible axial segments that is a strict superset of the other region; i.e:
\begin{align}
	\mathcal{N}_{i} \subset \mathcal{N}_j \text{ or } \mathcal{N}_j \subset \mathcal{N}_i.
\end{align}
Thus, we can define region $r_i$ to be ``higher access'' than $r_j$ if $\mathcal{N}_i \supset \mathcal{N}_j$, and ``lower access'' if the reverse holds.
In general, trajectories can bifurcate in going from the higher-access region to the lower-access region, but not vice-versa.

Now let us establish the boundary conditions for boundary $b_{ij}^a$ between region $r_i$ (in domain $d_a$) and $r_j$ (not in domain $d_a$).
We start by identifying the higher-access region. 
Then, taking the population for region $r_i$ in domain $d_a$ to be $p_i^m$, we start by finding the connected populations in region $r_j$. 
This set of populations will define a set $\mathcal{D}_{ij}^a$ of domains connected to domain $d_a$ at that boundary, i.e. the domains which contain those connected populations.
If $r_i$ is the higher-access region, then the boundary conditions now take the form:
\begin{align}
	f(\ve{Z})\bigr|_{b_{ij}^a,d_a} &\underbrace{=  f(\ve{Z})\bigr|_{b_{ji}^b,d_b} = \dots}_{d_b \in \mathcal{D}_{ij}^a}\\
	n_i \Gamma_Z^i \bigr|_{b_{ij}^a,d_a} &= \sum_{d_b \in \mathcal{D}_{ij}^a} n_i \Gamma_Z^i \bigr|_{b_{ji}^b,d_b}.
\end{align}
Note that if $r_j$ has no populations that are connected to $p_i^m$, then $\mathcal{D}_{ij}^a = \{\}$, and the boundary conditions reduce to the reflecting boundary conditions of an external boundary.

If, instead, $r_i$ is the lower-access region, then there will be only one population $p_j^n$ that is connected to $p_i^m$ in domain $d_a$.
Denote the domain associated with $p_j^n$ as $d_b$.
Then, we can establish the boundary conditions as in the previous paragraph, but starting with $p_j^n$ in $d_b$.
Explicitly, we first find all the populations $p_i^l$ in region $r_i$ connected to $p_j^n$, and thus construct the set of domains $\mathcal{D}_{ji}^b$ containing those connected populations.
Note that $\mathcal{D}_{ji}^b$ will, by definition, include $d_a$.
Then, the boundary conditions are:
\begin{align}
	f(\ve{Z})\bigr|_{b_{ji}^b,d_b} &\underbrace{=  f(\ve{Z})\bigr|_{b_{ij}^c,d_c} = \dots}_{d_c \in \mathcal{D}_{ji}^b}\\
	n_i \Gamma_Z^i \bigr|_{b_{ji}^b,d_b} &= \sum_{d_c \in \mathcal{D}_{ji}^b} n_i \Gamma_Z^i \bigr|_{b_{ji}^b,d_b}.
\end{align}

The above procedure defines all internal boundary conditions for the BAFP problem.

\begin{figure}
	\centering
	\includegraphics[width=0.7\linewidth]{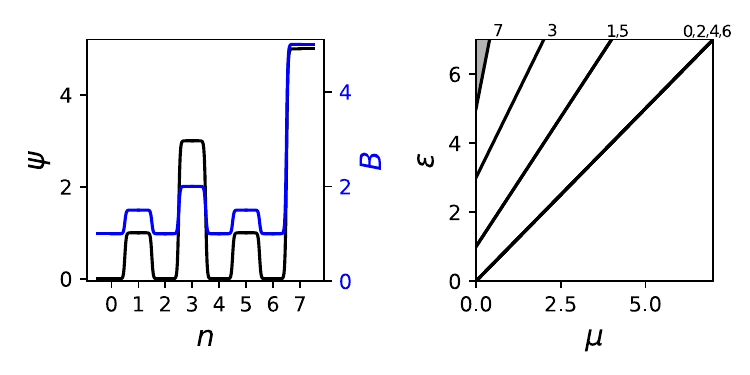}
	\caption{A ``tiered well'' field arrangement (left) and COM accessibility plot (right). In this scenario, in order of decreasing $\epsilon$, trajectories first bifurcate around axial segment $n=3$, then bifurcate again around segments $n=1$ and $n=5$. \label{fig:TieredScenarioSchem}}
	\includegraphics[width=\linewidth]{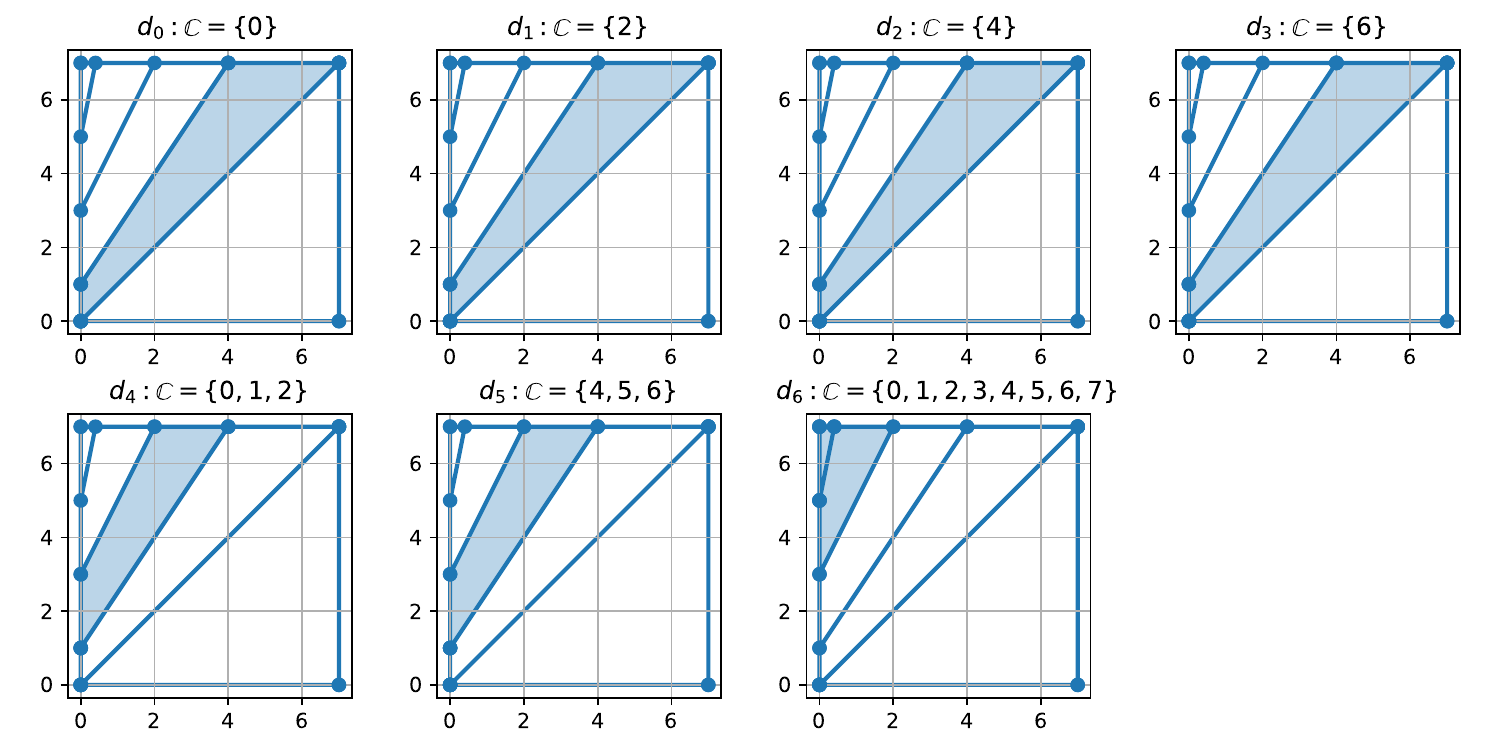}
	\caption{Domain decomposition for the ``tiered well'' in Fig.~\ref{fig:TieredScenarioSchem}.
		Each domain $d_a$consists of a list of populations, and as a result has an associated set of regions $\mathbb{R}_a$ (the shaded area in each plot), and continuous set of axial segments $\mathbb{C}_a$, given in the title of the plot. Here, several domains (e.g. $d_0$, $d_1$, $d_2$, and $d_3$) share the same region in $(\epsilon,\mu)$, but represent different populations because they occur at different positions along the field line. \label{fig:TieredScenarioDomains}}
	\includegraphics[width=0.5\linewidth]{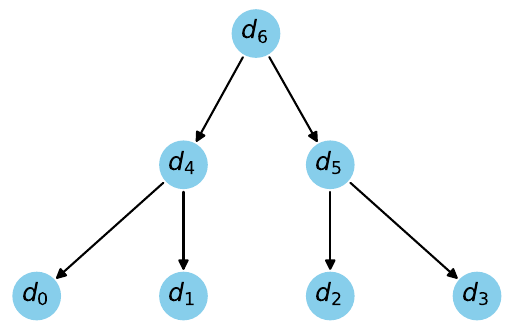}
	\caption{Graph structure of domain connections for the ``tiered well'' in Fig.~\ref{fig:TieredScenarioSchem}, with the domains defined in Fig.~\ref{fig:TieredScenarioDomains}.
		The forks in the graph are associated with the trajectory bifurcations, first at $n=3$ (splitting $d_6$ into $d_4$ and $d_5$), then again around segments $n=1$ and $n=5$.\label{fig:TieredScenarioGraph}}
\end{figure}

\section{Graph Structure of Domain Connections} \label{sec:Graphs}

The rules outlined in Secs.~\ref{sec:DomainRules}-\ref{sec:BoundaryConditions} are sufficient to properly set up a BAFP problem, to solve it either analytically or (more likely if application is necessary) computationally.
However, it is also useful when thinking about a problem to be able to quickly visualize how the various domains fit together.
The rules above, which are focused on conditions at each boundary (of which each domain can have quite a number), do not make developing this intuition particularly straightforward.
The goal, then, is to quickly encode the boundary condition connections into the connections between domains, which can then be visualized.

The condition for two domains to be connected is simple: domains $d_a$ and $d_b$ will be connected if
\begin{align}
	\exists p_i^m \in d_a, p_j^n \in d_b: (p_i^m \text{ and } p_j^n \text{ are connected.})
\end{align}
Furthermore, we can establish a ``direction of connection,'' based on whether $p_i^m$ or $p_j^n$ is in a higher-access region.
We draw the direction from the higher-access region to the lower-access region, representing the direction in which trajectories can bifurcate. 
In other words, if $\mathcal{N}_i \supset \mathcal{N}_j$, the direction of connection is from $d_a$ to $d_b$.
Note that if two domains connect at multiple boundaries, the direction of connection can in principle be bi-directional.

With such connections established, the boundary condition structure of a BAFP problem can then be represented as a graph.
This graph can help to clarify the multi-well structure of the BAFP problem, in particular encoding the bifurcation structure of trajectories.

For instance, Fig.~\ref{fig:TieredScenarioSchem} shows a ``tiered well,'' where as $\epsilon$ decreases, trajectories first bifurcate at the local potential and magnetic maximum at $n=3$, and then bifurcate again at the local potential and magnetic maxima at $n=1$ and $n=3$.
Although there are only three resulting regions in $(\epsilon,\mu)$ space, the separate wells mean that there are 7 separate domains (Fig.~\ref{fig:TieredScenarioDomains}).
The boundaries between these domains are connected as shown in Fig.~\ref{fig:TieredScenarioGraph}.
Domain $d_6$ represents the domain of trajectories which transit the entire device.
Then, this bifurcates into domains $d_4$ and $d_5$, which represent the trajectories trapped on either side of $n=3$, but which pass over the potential maxima at $n=1$ and $n=5$.
Finally, the trajectories bifurcate again at these final potential maxima.
In this way, the graph structure succinctly encodes the domain connections that result from the well structure.

\begin{figure}
	\centering
	\includegraphics[width=0.7\linewidth]{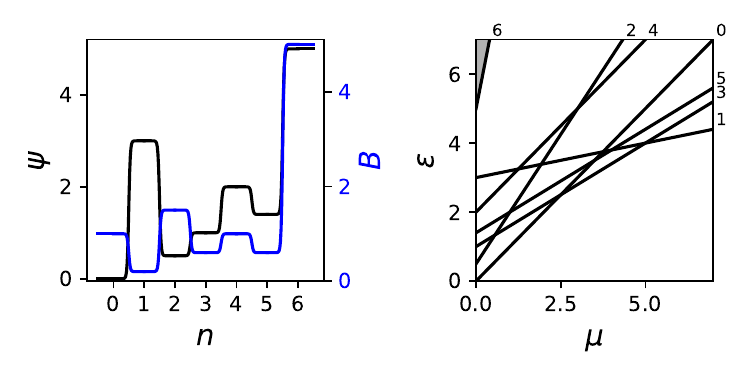}
	\caption{An arbitrary complicated field arrangement, with many crossings in the accessibility boundary lines. \label{fig:ComplicatedScenarioSchem}}
	\includegraphics[width=\linewidth]{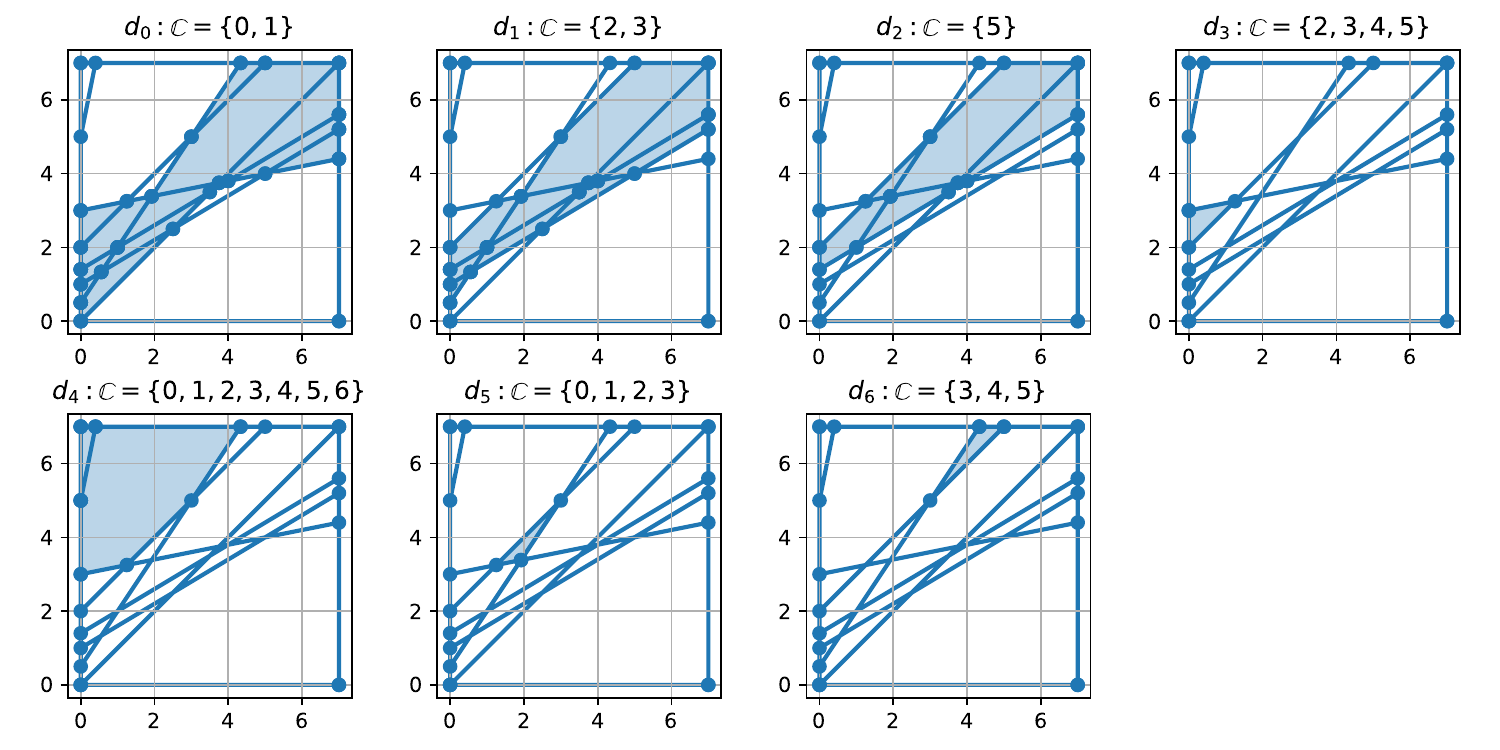}
	\caption{An algorithmically-solved consistent domain decomposition of the scenario in Fig.~\ref{fig:ComplicatedScenarioSchem}, showing the shaded set of regions $\mathbb{R}_a$ and continuous set of axial segments $\mathbb{C}_a$ for each domain $d_a$. \label{fig:ComplicatedScenarioDomains}}
	\includegraphics[width=0.5\linewidth]{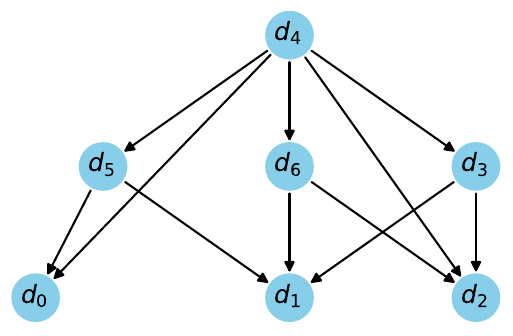}
	\caption{Graph structure of domain connections for the scenario in Fig.~\ref{fig:ComplicatedScenarioSchem}, with the domains defined in Fig.~\ref{fig:ComplicatedScenarioDomains}.
		The graph structure is much more complicated than for the tiered well scenario in Figs.~\ref{fig:TieredScenarioSchem}-\ref{fig:TieredScenarioGraph}, but if one chooses any set of connected nodes it is possible to identify the boundary where the conditions are enforced.\label{fig:ComplicatedScenarioGraph}}
\end{figure}

Of course, the ``tiered well'' is very straightforward example with a very clear logic to the domain connections.
The established rules for establishing cohesive domains and determining the connections between domains really prove their value when considering more complicated scenarios.
For instance, consider the field arrangement in Fig.~\ref{fig:ComplicatedScenarioSchem}.
One might scratch their head for quite a while coming up with a set of consistent domains to describe this problem, but the rules in Sec.~\ref{sec:DomainRules}, as implemented in the algorithm in Appendix~\ref{sec:Algorithm}, can quickly come up with a set of consistent domains (Fig.~\ref{fig:ComplicatedScenarioDomains}).
Furthermore, the methods described here can also quickly reveal the connections between these domains (Fig.~\ref{fig:ComplicatedScenarioGraph}).
Thus, it can be seen that formulating the problem in this way opens up the possibility of solving arbitrarily complicated multi-well BAFP problems.

\section{Conclusion} \label{sec:Conclusion}

In this paper, we have shown how to construct consistent domains for the solution of BAFP problems, so that in each domain, a single point in COM space corresponds to a single trajectory.
We also showed which boundary conditions were enforced between the resulting domains, and how these boundary connections (associated with trajectory bifurcations) could be succinctly summarized via directed graphs.

Though the focus was primarily on magnetic mirrors, the generality of the derived conditions should allow bounce-averaged theories to be applied to a variety of plasma systems with well-defined COM, including tokamaks and quasisymmetric stellarators.
Crucially, the clear rules for identifying consistent domains should enable the gridding and boundary-matching process to be automated, allowing bounce-averaged theories to be applied even in systems where wells (and their associated trajectory bifurcations) arise dynamically and unpredictably over the course of the simulation.
This is the case, for instance, in the simulation of magnetic mirrors with highly kinetic sloshing ion distrbutions, or arrangements in which Yushmanov trapping becomes significant. 
Of course, there are other issues to be solved in making bounce-averaged simulations work efficiently for such plasmas, including the development of stable and efficient methods for the ambipolar solve discussed in Sec.~\ref{sec:AmbipolarSolve}, as well as efficient methods for choosing between the various possible domain decompositions so as to retain maximal efficiency and continuity.
Nevertheless, the present work provides a crucial stepping stone towards enabling BAFP codes to simulate these more complicated and relevant modern systems, providing a self-consistent alternative to much more expensive codes that resolve the parallel dynamics.

\section*{Acknowledgements}

The author thanks Nat Fisch, Alex Glasser, Thomas Foster, and Elijah Kolmes for valuable conversations.

\section*{Funding}
This work was supported by ARPA-E Grant No. DE-AR0001554. 

\section*{Declaration of Interests}
The authors report no conflict of interest.

\section*{Author ORCID}  
Ian E. Ochs, https://orcid.org/0000-0002-6002-9169.



\appendix

\section{Differential Condition for Yushmanov Trajectory Bifurcation} \label{sec:TrajectoryBifurcationNonRelativistic}

Here, we derive the nonrelativistic condition for a trajectory bifurcation to occur when the potential is increasing and the magnetic field is decreasing as a function of axial coordinate $s$ (or vice-versa); i.e. when 
\begin{align}
	\text{sgn}\left(\pa{B}{s}\right) = -\text{sgn}\left(\pa{\psi}{s}\right). \label{eq:YushmanovBifurcationPrerequisite}
\end{align}
We call this a ``Yushmanov trajectory bifurcation'' because it is closely related to the off-midplane ``Yushmanov-trapped'' ions that are well-known from mirror theory \citep{Post1987MagneticMirror,Yushmanov1966ConfinementSlow}.

Thus, recall Fig.~\ref{fig:COMPlot_BifurcationNoMax}.
The reason a trajectory bifurcation occurs in this case is that lines 0 and 2 intersect below line 1.
We can easily write this condition in terms of $B_n$ and $\psi_n$.
The intersection point $\mu^*$ of lines 0 and 2 is given by:
\begin{align}
	\mu^* B_0 + \psi_0 = \mu^* B_2 + \psi_2 \Rightarrow \mu^* = \frac{\psi_2 - \psi_0}{B_0 - B_2}.
\end{align}
Then, the value of the accessibility boundary at this $\mu^*$ must be lower for line 0 than for line 1:
\begin{align}
	\mu^* B_0 + \psi_0 < \mu^* B_1 + \psi_1. \label{eq:YushmanovBifurcation1}
\end{align}

We can turn the condition into a differential condition by taking the axial segments to be equally spaced with spacing $\Delta s$, and taking $\Delta s \rightarrow 0$. 
This requires working to second order in $\Delta s$:
\begin{align}
	\psi_0 &= \psi_0; \\
	\psi_1 &= \psi_0 + \pa{\psi}{s} \Delta s + \frac{1}{2} \pa{^2\psi}{s^2} (\Delta s)^2;\\
	\psi_2 &= \psi_0 + \pa{\psi}{s} (2 \Delta s) + \pa{^2\psi}{s^2} (2 \Delta s)^2.
\end{align}
The same expansion is enforced for $B$.
With this prescription, Eq.~(\ref{eq:YushmanovBifurcation1}) becomes:
\begin{align}
	\frac{\paf{^2B}{s^2}}{\paf{B}{s}} > \frac{\paf{^2\psi}{s^2}}{\paf{\psi}{s}} . \label{eq:YushmanovBifurcation2}
\end{align}
Thus, if Eqs.~(\ref{eq:YushmanovBifurcationPrerequisite}) and (\ref{eq:YushmanovBifurcation2}) hold at a point, there will be a trajectory bifurcation there.

\begin{figure*}
	\centering
	\includegraphics[width=\linewidth]{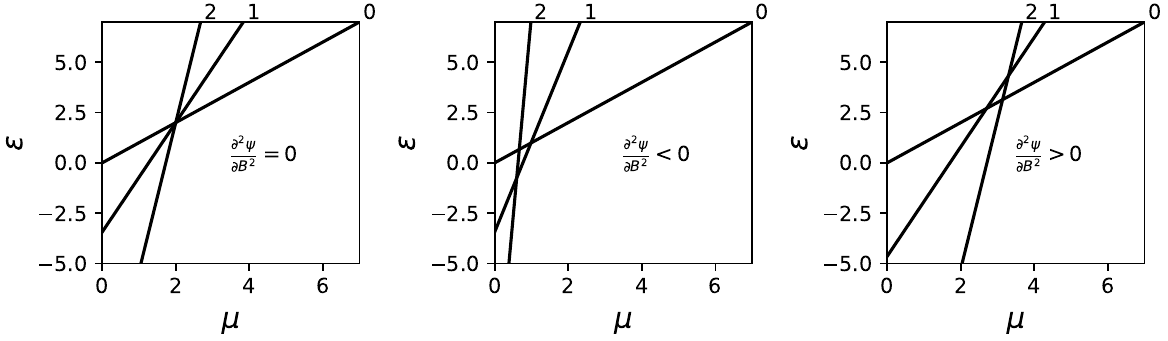}
	\caption{Accessibility region boundaries in COM space for $s=0,1,2$ for fields of the form in Eq.~(\ref{eq:YushmanovExampleField}), demonstrating the Yushmanov trajectory bifurcation condition [(Eqs.~(\ref{eq:YushmanovBifurcationPrerequisite}) and (\ref{eq:YushmanovBifurcation2}) or Eqs.~(\ref{eq:YushmanovBifurcationNoCoordinates})]. 
		In the first panel, $a_B = a_\psi =1$, and the lines all intersect at a single point, so there is (marginally) no bifurcation. 
		In the second panel, $a_B = 1 < a_\psi = 1.2$, and so there is no bifurcation.
		In the third panel, $a_B = 1.5 > a_\psi = 1$, and so Eq.~(\ref{eq:YushmanovBifurcation2}) [or Eq.~(\ref{eq:YushmanovBifurcationNoCoordinates})] is satisfied and there is a trajectory bifurcation. 
		For all plots, $c_B = 1$ and $c_\psi = 2$. \label{fig:YushmanovBifurcation}}
\end{figure*}

Alternatively, conditions~(\ref{eq:YushmanovBifurcationPrerequisite}) and (\ref{eq:YushmanovBifurcation2}) can also be put into $s$-independent form by treating $\psi$ as a function of $B$ and expanding $\psi(B)$, yielding the equivalent but simpler conditions for a trajectory bifurcation:
\begin{align}
	\pa{\psi}{B} < 0 \text{ and } \pa{^2\psi}{B^2} > 0. \label{eq:YushmanovBifurcationNoCoordinates}
\end{align}

To see these conditions in action, we can take fields of the form:
\begin{align}
	B = c_B e^{a_B s}; \; \psi = c_\psi  (1-e^{a_\psi s}), \label{eq:YushmanovExampleField}
\end{align}
with $a_B >0$ and $a_\psi >0$ to satisfy Eq.~\ref{eq:YushmanovBifurcationPrerequisite}.
Then, Eqs.~(\ref{eq:YushmanovBifurcation2}) and (\ref{eq:YushmanovBifurcationNoCoordinates}) reduce to the condition that $a_B > a_\psi$.

The accessibility boundaries for fields of the form of Eq.~\ref{eq:YushmanovExampleField} are shown in Fig.~\ref{fig:YushmanovBifurcation} for $s\in \{0,1,2\}$.
It can be seen that the condition $a_B > a_\psi$, corresponding to Eq.~(\ref{eq:YushmanovBifurcation2}), correctly describes the onset of the bifurcation.

\section{Non-Transitivity of Compatibility and Non-Uniqueness of Domain Decomposition} \label{sec:Nontransitivity}

In this appendix, we show that the compatibility conditions are not necessarily transitive, and that the domain decomposition is not necessarily unique.
To show this, an example field configuration suffices.

\begin{figure}
	\centering
	\includegraphics[width=0.8\linewidth]{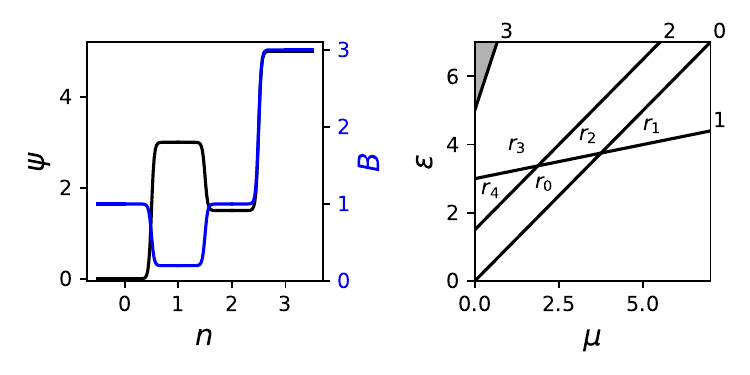}
	\caption{Example field configuration and COM-space accessibility plot demonstrating non-transitivity of population compatibility and non-uniqueness of the domain decomposition. \label{fig:COMPlot_NonTransitivity}}
\end{figure}

Thus, consider the field arrangement in Fig.~\ref{fig:COMPlot_NonTransitivity}.
This grid is characterized by 6 populations:
\begin{align}
	p_0^0&: \{r_0,\mathcal{C}_0^0 = \{0\}\}; \notag\\
	p_1^0&: \{r_1,\mathcal{C}_1^0 = \{1\}\};\notag\\
	p_2^0&: \{r_2,\mathcal{C}_2^0 = \{0,1\}\};\notag\\
	p_3^0&: \{r_2,\mathcal{C}_3^0 = \{0,1,2\}\};\notag\\
	p_4^0&: \{r_4,\mathcal{C}_4^0 = \{0\}\};\notag\\
	p_4^1&: \{r_4,\mathcal{C}_4^1 = \{2\}\}.\notag
\end{align}

There is a clear incompatibility between the populations $p_3^0$ in region $r_3$, and the populations $p_4^0$ and $p_4^1$ in region $r_4$.
However, every other population is compatible with both population $p_3^0$ and $p_4^0$.
Thus, the compatibility condition is seen to be non-transitive.

This non-transitivity also leads to non-uniqueness of the domain set.
Due to the bifurcation in trajectories between regions $r_3$ and $r_4$, 3 domains are required.
However, which populations go into these domains is a matter of choice.
For instance, we can simply assign each of the troublesome bifurcated populations in region $r_4$ to their own domain:
\begin{align}
	d_0 = \{p_0^0,p_1^0,p_2^0,p_3^0\}; \; d_1 = \{p_4^0\}; \; d_2 = \{p_4^1\}.
\end{align}
Alternatively, however, we can also isolate population $p_3^0$ instead of $p_4^0$.
Then we get:
\begin{align}
	d_0' = \{p_0^0,p_1^0,p_2^0,p_4^0\}; \; d_1' = \{p_3^0\}; \; d_2' = \{p_4^1\}.
\end{align}
Either of the domain decompositions are equally valid according to the conditions laid out in Sec.~\ref{sec:DomainRules}, and thus the decomposition is not generally unique.

\section{Algorithm to Find a Complete Set of Domains} \label{sec:Algorithm}

A complete partitioning of the populations into domains can be performed as follows.
\begin{enumerate}
	\item Define $\mathcal{U} = \{p_i^m\}$, the set of unsorted populations.
	\item Define $\mathcal{D} = \{\}$, the set of domains.
	\item If $|\mathcal{U}| > 0$, initialize a new domain $d = \{\}$; otherwise terminate. \label{algo:DomainInitialize}
	\item Remove a population $p_i^m$ from $\mathcal{U}$, add it to $d$.
	\item Construct $\mathcal{R}_i$, the set of neighboring regions of region $r_i$, where $r_i$ is the region associated with populations $p_i^m$. \label{algo:PopulationLoopBegin}
	\item Choose an $r_j \in \mathcal{R}_i$. \label{algo:RegionSelect}
	\item Check if a population from $r_j$ is already in $d$; if yes, go to the next region in $\mathcal{R}_i$.
	\item Check if a population $p_j^n \in \mathcal{U}$ with region $r_j$ is connected to population $p_i^m$ (Condition~\ref{eq:ConnectedPopulations}) and compatible with all populations in $d$ (Conditions~\ref{eq:Compatiblity1}-\ref{eq:Compatiblity2}). \label{algo:ConnectedCompatibleCheck}
	\item If \ref{algo:ConnectedCompatibleCheck}, remove $p_j^n$ from $\mathcal{U}$, and add $p_j^n$ to $d$. \label{algo:PopulationAdd}
	\item Repeat \ref{algo:RegionSelect}-\ref{algo:PopulationAdd} until all regions in $\mathcal{R}_i$ have been checked. \label{algo:PopulationLoopEnd}
	\item Repeat \ref{algo:PopulationLoopBegin}-\ref{algo:PopulationLoopEnd} for each population $p_i^m$ in $d$.
	\item When all neighboring regions of all populations have been checked, and no new populations can be added, add $d$ to $\mathcal{D}$, and go to \ref{algo:DomainInitialize}.
\end{enumerate}

It is also useful to have a computer-friendly algorithm for step \ref{algo:ConnectedCompatibleCheck} to determine compatibility and connectedness.
To determine whether $p_i^m$ has a compatible population in region $r_j$, one can do the following:
\begin{enumerate}
	\item Count the number $a$ of elements $\mathcal{C}_j^n$ of $\boldsymbol{\mathcal{C}}_j$ s.t. $\mathcal{C}_j^n \cap \mathcal{C}_i^m \neq \{\}$.
	\item If $a = 1$, then choose the single $\mathcal{C}_j^n$ that intersects $\mathcal{C}_i^m$. 
	\item Count the number $b$ of elements $C_i^k$ of $\boldsymbol{\mathcal{C}}_i$ s.t. $\mathcal{C}_i^k \cap \mathcal{C}_j^m \neq \{\}$.
	\item Then:
	\begin{enumerate}
		\item If $a = 0$, $p_i^m$ is compatible with all populations in $r_j$, but not connected.
		\item If $a > 1$, then $p_i^m$ is connected to multiple populations in $r_j$, so there is no compatible population in $r_j$ (there is a trajectory bifurcation going from $r_i$ to $r_j$).
		\item If $a=1$ but $b>1$, then $p^n_j$ is connected to multiple populations in $r_i$, so $p_j^n$ and $p_i^m$ are not compatible (there is a trajectory bifurcation going from $r_j$ to $r_i$).
		\item If $a = 1$ and $b=1$, then $p_i^m$ is compatible with $p_j^n$ in region $r_j$, with $C_j^m$ given by step 2. If $r_i$ borders $r_j$, these populations are also connected.
	\end{enumerate}
\end{enumerate}
If one wants to check whether a specific population $p_j^l$ in region $r_j$ is compatible with $p_i^m$, then one must additionally check that the compatible population $p_j^n$ identified in step 4d is the same as the queried population $p_j^l$.


\clearpage
\newpage

\end{document}